# Hubs in Languages' Scale Free Networks of Synonyms

Hanna E. Makaruk
P-22, MS D454, Los Alamos National Laboratory,
Los Alamos, NM, 87-545, USA
hanna_m@lanl.gov

Robert Owczarek,
EES-11, MS D443, Los Alamos National Laboratory,
Los Alamos, NM, 87-545, USA
rom@lanl.gov

Natural languages are described in this paper in terms of networks of synonyms: a word is identified with a node, and synonyms are connected by undirected links. Our statistical analysis of the network of synonyms in Polish language showed it is scale-free; similar to what is known for English. The statistical properties of the networks are also similar. Thus, the statistical aspects of the networks are good candidates for culture independent elements of human language. We hypothesize that optimization for robustness and efficiency is responsible for this universality.

Despite the statistical similarity, there is no one-to-one mapping between networks of these two languages. Although many hubs in Polish are translated into similarly highly connected hubs in English, there are also hubs specific to one of these languages only: a single word in one language is equivalent to many different and disconnected words in the other, in accordance with the Whorf hypothesis about language relativity. Identifying language-specific hubs is vitally important for automatic translation, and for understanding contextual, culturally related messages that are frequently missed or twisted in a naïve, literary translation.

*Keywords:* synonyms, scale free network

## 1. Introduction

The science of networks has become an important branch of studies of complex systems (systems with very large or infinite number of degrees of freedom). A good introduction to networks is provided by Barabasi [4]. More advanced approaches are found, for example, in [1, 6]. In order to make this paper somewhat self-contained for a reader less familiar with these concepts, we include a short Appendix, in which some definitions and facts are listed. Here we state somewhat informally that systems of objects may be described as networks (graphs), with the objects usually identified with vertices (nodes) of the graph, and interactions of objects identified with edges of the graph. This simplification ignores details of interactions, focusing on the global features of a system.

The science of networks turned out to be fruitful in the description of natural languages. In particular, the structure of the network of synonyms in English was studied by Motter et al. [20] (also, e.g. by Holanda et. al. [14] but the latter studies are difficult to compare with ours). In the network, words were considered as nodes. Synonyms were connected by undirected link. Statistical properties of the resulting network of synonyms in English showed that it is scale-free. Similarly, in the Italian language [1, 2] various statistical characteristics, including frequency and entropy of punctuation marks, lengths of words, were show to follow the functional dependence predicted by Pareto's law [21]. Other studies [10, 11, 12], focused on syntactic connections between words in sentences, and found that they created a scale-free network, too.

In this paper, the Polish language is considered in a framework similar to that used by Matter at al. for English [20]. The network of synonyms in Polish is found to be a scale-free network as well. The statistical properties of both languages are very similar. We have also analyzed the problem of one-to-one (at least in some approximate sense) mapping of the hubs i.e. nodes, particularly those with many connections, to similarly highly connected hubs between English and Polish. Some of them can be mapped, onto each other, but there are also a number of hubs, which cannot. As a result, a unique mapping of one network to the other does not exist. This is an important fact for consideration in any attempt at automatic translation. The language specific hubs contribute also, as discussed later, to contextual meaning of a message, which is so frequently lost in a naïve "first meaning only" translation.

In the first section, the data gathering method is discussed, and the results of the statistical analysis for the Polish language are presented. In the second section, the phenomenon of language-specific synonyms is presented and discussed in the light of the Whorf hypothesis of linguistic relativity, which connects perception of the world to certain aspects of the native language, and has consequences discussed in many linguistic papers, e.g. [19]. In the summary, open questions and the possibility of application of the results is discussed. Definitions and facts used in the mathematical part of this paper are listed in the Appendix.

## 2. Network of Synonyms in Polish Language

In order to study the network of synonyms in Polish language, we had to decide first on a source. We used as a source the dictionary [17] with the content of over 23000 words and expressions. To allow for comparison between languages, our methodology for this study follows the methodology described in [20], as closely as possible. The data were prepared as follows:
  a. The dictionary is treated as a model of a living language.
  b. All the words included in one meaning of the main word are treated as connected, but words included in two different meanings of the same hub- word are not.
  c. The words and expressions included in the dictionary were not changed in any way except by stripping phonetic accents from them. It is assumed that words without phonetic accents (such phonetic accents are very common in Polish, so it is an important issue) are still uniquely identified. The exceptions from this rule are so rare that the issue is not treated as statistically significant here. A few of them are *bak, chrzest, czesc, zadanie, zle*.
  d. For words with a huge number of synonyms, dictionaries establish a pattern, rather than attempt to list all of them. Example: *Drink beverage, tea, coffee, milk…* It introduces systematic error, similar for all of the synonyms dictionaries. The problem has been noticed in the literature [20].

After the preparation of the data as described above, the statistical properties of the network of synonyms were studied. The dictionary was scanned; scanned images of the pages converted into a text file, and then the text words converted into variables in Mathematica program in which all the calculations were made. The results are shown in the two figures below.

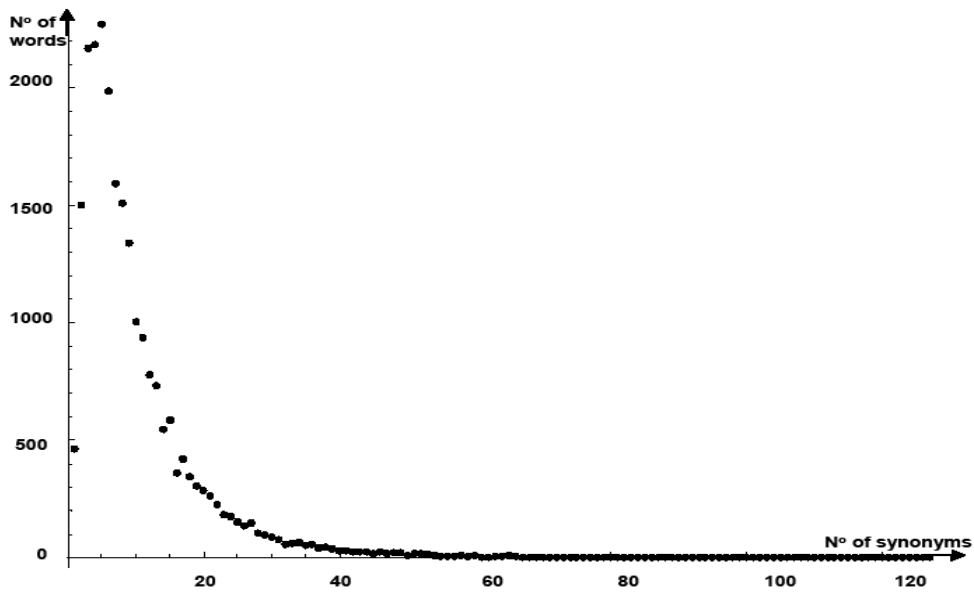

Fig.1 The number of synonyms vs. the number of words in Polish (linguistic data from [17])

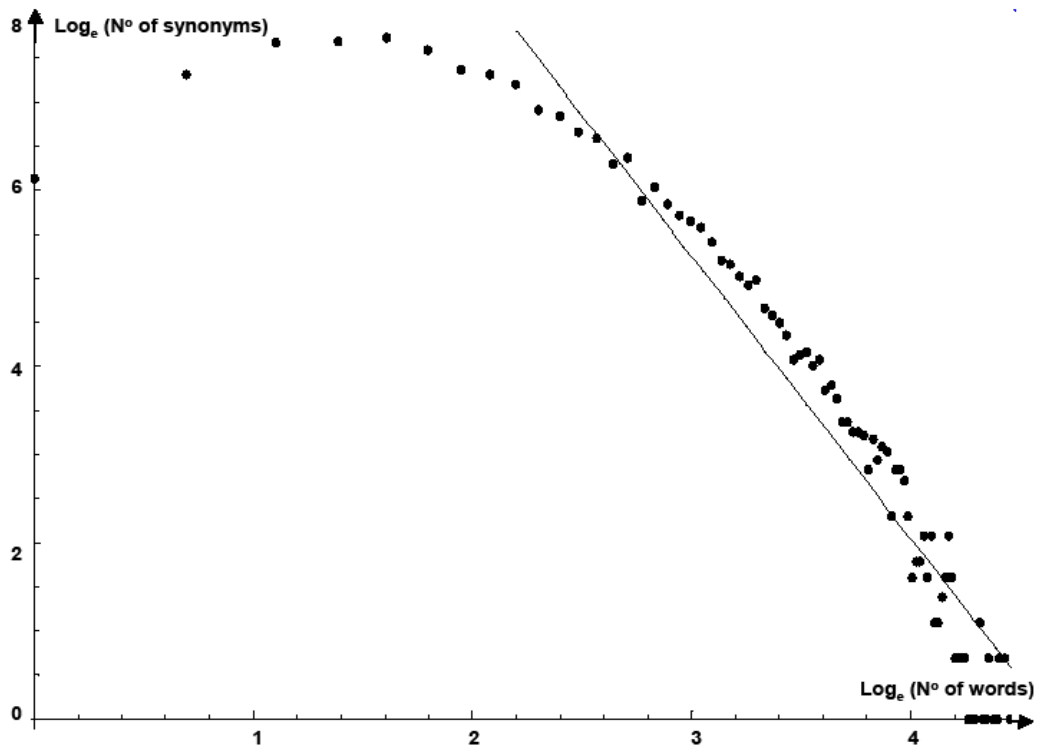

Fig.2 Data from the Fig.1 with both scales logarithmic. Due to systematic error discussed in (d) the data fulfill the scale free low – approximates the straight line only in some range of the number of synonyms.

The basic result is that the network of synonyms in Polish is a scale-free network with $\gamma \approx 3.20$ (see the definition in the Appendix). If we compare this result with the results for the network of synonyms in English [20], we can conclude that in this statistical sense Polish is very similar to English ($\gamma \approx 3.5$), because the dependence is power-like and the values of the power exponent $\gamma$ are very close in these two languages. In addition, we can conclude that there exists a threshold for disconnection.

As mentioned before, no dictionary of synonyms is a perfect representation of a living language, but in case that perfect mapping of a living language structure into a dictionary would be possible, $\gamma$ is still predicted to be over 3. Existence of a threshold in the number of words is known to everyone who learned any foreign language knows that there is a threshold in the number of known words. Until the mastered vocabulary is larger than the threshold (typically a few hundreds of words), it is next to impossible to communicate in real life. At such stage, despite knowledge of proper grammar, one is not able to connect known words into meaningful sentences, except for a few artificially prepared ones to be found in a language textbook; the language network is disconnected. It shows existence of the threshold.

Another observation is that "old" words have more synonyms. Namely, among words in Polish with 1 synonym 76% are "new" words, while among hubs with 45 to 88 synonyms only 18% are "new" words. The distinction between "old" and "new" words may of course be disputable in a few individual cases, but for the vast majority of the words it is easy to make. The two criteria applied are the meaning and the root of a word. A word is treated as "old" if it was used in the eighteenth century, and as "new" if it is was brought into Polish language during the nineteenth-century industrial revolution or later. The "old" words have almost exclusively Slavic or Latin roots (a small number of the "old" words have German or, very rarely, other roots). The "new" words were brought from other languages or were artificially developed from Slavic roots to describe new concepts introduced by the industrial revolution. Examples of words treated as new: *robot* (from Czech (and Polish) *robota*= work, labor), *komputer* (from computer in English).

We hypothesize why the language developed as a scale-free network. Languages survived centuries of cultural changes; they had to be easily understandable when spoken, even in high-noise environments, which cause random corruption of a message. Languages had to be functional also when vocabulary was limited, which means, only a small fraction of nodes were in use, since otherwise children would not be able to communicate and effectively learn a language. Obviously the same concerns foreigners, nevertheless children are the crucial factor for language preservation. As a result, any natural language is optimized for robust communication. Scale-free structure allows for such optimization. Natural languages can serve as very good examples of robust, scale free networks, whose mathematical properties could be imitated in artificial networks aiming for robustness like high level programming languages, Internet connections, and network of airports.

Natural languages have a level of redundancy sufficient to provide efficient communication in unfavorable conditions (noise) and to allow for communication between subjects with limited language skills- small children and foreigners can communicate while knowing under 10% of the total vocabulary of a particular language [15]. It is the very property of the language, which allows for learning it. The redundancy is on the other hand limited by the information-entropy value. Humans in general do not want to devote excessive words, time, energy to describe a thing or a concept if there exists simpler yet fully efficient way of doing it. By analogy, no one wants to invest in costly overdevelopment of Internet servers and connections, additional airports etc, if a smaller network is sufficient and functional even when some nodes are randomly out of use for a variety of reasons. Overcomplication of a programming language is also costly; it extends computation time and effectively limits the size of a program, which could be run on a particular hardware. The overdevelopment of airports structures, which do not contribute to the air traffic volume, would be costly. This is why information about mathematical and statistical properties of scale free networks, which were optimized over centuries for efficiency and robustness, should be applied directly in the design of artificial scale free networks. This is also, why it is crucial to distinguish universal properties of natural languages from culture/language specific ones. Language specific properties were developed due to unique cultural and historical reasons, unrelated to the network robustness. The universal properties are connected with optimization of the network design, the culture/language specific properties are important in contextual interpretation of messages and translation between languages.

Many scientists theorize that all languages are built in the same way, in attempt to describe the universal human experience [20]. Others concentrate on the Whorf hypothesis of language relativity, pointing to the fact that even such simple notions as names of colors cannot be mapped bijectively (one-to-one) between languages and the whole concept of two colors being the same or different depends on the native language of the person judging it [8].

An overall framework is needed, clearly distinguishing between universal and culture specific properties of languages. Theorizing about this based on any individual language is premature [20], we show examples, which contradict some generalizations made in [20]. Comparison between English (Germanic language with strong Latin influence) and Polish (Slavonic) allows us also to support the major thesis of [20] about scale free network character of natural languages and show similarities between these two networks of synonyms; nevertheless there is no chance for a bijective mapping between these two networks. There exist hubs, each of which has a counterpart in the other language that is also a hub, as predicted by Motter et al. [20]. On the other hand, contradictory to [20], there exist also some hubs (examples of them are discussed in the next section), which appear in only one of these languages and totally change local topology of the network.

## 3. Language-specific hubs in networks of synonyms

Despite the similarity in statistical sense, there is no simple map of the network of synonyms for Polish onto the analogous network for English. The reason is found when individual hubs in the networks are studied.

Let us focus on words, which have sets of meanings unique to a particular language. In other languages these meanings are not connected.
Examples of synonyms in English, for which meanings are not connected in Polish:
- *Blue* (1) color, (2) mood
- *Drum* (1) musical instrument, (2) container

- *Charge* (1) electrical charge, (2) basis for arrest

Examples of synonyms in Polish, for which meanings are not connected in English:
- *Niebieski* (adjective) (1) astronomical (2) heavenly, (3) blue color, (4) connected with the sky
- *Pokój* (1) peace, (2) room in a house
- *Zachód* (1) West-direction; (2) sunset-time of the day; (3) moment when a star or astronomical object vanishes under the horizon; (4) developed countries (5) NATO (6) time consuming, tearing, non-productive task, leg-work.
- *Południowy*- adjective derived from (1) lunch time, (2) South-direction (3) expressive, emotional (about personality) (4)extremely hot (about whether ) (5) dark (about pigmentation)
- *Pierwiastek* (1) cultural: influence, (2) chemical: element, (3) mathematical: root
- Rola (1) actor's role, (2) agricultural field
- Ziemia (1) planet Earth, (2) soil, (3) country

Color recognition and distinction of basic colors in different languages is commonly recognized [8] as a good test field for the Whorf hypothesis about language relativity [19]. In Polish two colors are used to describe things described in English as white. Both of them differ from silver:
*Biały* is a covering color. It is a color of paper, chock, and a white wall.
*Siwy* express colorlessness. It describes human hair turned white, similar color of animal fur, dense fog, light shadows of a smog.
There is no ambiguity in mind of a native Polish speaker, when asked, to differentiate between these two, different colors.

The interesting consequence of the above mentioned differences are observed cultural differences related to the language in our understanding of the world. An example of this phenomenon might be the difference in understanding of the concept of time in English and in Mandarin, as explained by Boroditsky in [5]. The English speakers are used to understand time as a horizontal line, while in Mandarin it is equally accepted to think of it in terms of a vertical line, which leads in consequence to different sense of time in the two cultures. However, the connection between culture and language has been challenged, for example, by Hadley [13]. Nevertheless, it seems that cultural background adds another layer of difficulty in learning a new language [7, 9]. Another curious issue is so-called personality switch characteristic for bilinguals, who seem to be culturally closer to the language they are currently using [22] (more on similar subjects in [16, 18].)

Language specific hubs seem to be overlooked in linguistics. Textbooks and teachers of foreign languages are routinely silent about the subject, leaving the students the task of finding by trial and error that some notions, which are identical or very similar for them, are considered completely disconnected in the language they learn. Software for automatic language translation is known to neglect the topic as well, which sometimes results in funny mistranslations, but more frequently in completely disorganized and confusing text. Taking into account language specific synonyms would certainly improve quality of such software, and make the lives of language students much easier.

Each message consists of two elements: literal meaning and a context. Concentration on literal meaning without ensuring that the context did not change makes no more sense than consideration of a mathematical formula for a function without regard to its domain. Cultural context consists of multiple elements, not all of them are built into the language, but one, quite important, is. When a word is having multiple separate meanings, it can be used for one meaning but giving also a hint about the other one. In scientific and everyday communication people aim for precision and tend to avoid words with multiple or unclear meaning. Such texts rarely contain translation traps. Ambiguity, multiple meanings, relations between the meanings, emotions are explored in some poetry and jokes, both of them notoriously difficult to translate. These two areas may sound marginal, but there is the third, very important area – communication by open channels between people connected with an illegal or semi legal political movement, terrorist group etc. It should not be mixed with cryptography – like communication in an organized gang or other criminal group where message is sent to a well-defined receiver who knows a kind of internal code of the group, preestablished more or less formally before the message is sent. Political groups aim for broad audiences, and obviously have no way to provide any kind of distribution of the key to their encrypted messages. The messages are frequently hidden between thousands of neutral news items in a newspaper, TV program, etc., so the receiver has to start by identifying existence of the message, which has escaped censorship, and where it begins and where it ends. Words with many different meanings are typically used there.

## 4. Summary

Natural human languages are networks, which were optimized for both robustness and efficiency during their evolution. If the network characteristics, which are responsible for this robustness, are properly identified, they may be applied in the construction of many kinds of artificial networks. These characteristics are expected to be culture independent, in contrast to language specific characteristics, which the most probably evolved in one language (a family of languages) just by chance, or within a particular cultural context. In looking for these language independent characteristics, the structure of synonyms in Polish language was compared to the structure of synonyms in English. Both languages have striking similarities, from the statistical point of view. They can be considered scale free networks. Moreover, the statistical coefficients are found to be similar: $\gamma \approx 3.2$ vs. $\gamma \approx 3.5$. This statistical structure allows for language robustness Major hubs-general words, which are synonyms to many specific ones, are the same. Nevertheless, the networks cannot be considered as maps of each other as there are a considerable number of words, which are hubs in only one of the languages. Different meanings of such word are synonyms in one language and are not connected at all in the other. They are strong illustrations of the Whorf hypothesis of linguistic relativity. This also means that there is no possibility for constructing a one-to one mapping between the synonyms networks of two different languages.

# 5. Acknowledgment

Authors thank Joysree Aubrey for the in-depth discussions and suggestions about the contextual culturally related messages. Authors also thank Józef Przytycki for the discussion about the mathematical properties of language as a scale free network.

# 6. Appendix

**Definition**: A graph is a pair (V, E), where V is a set of points, called nodes or vertices, and E is a set of (unordered) pairs of points from V.

**Remarks**: For the purposes of this paper, we do not distinguish between networks and graphs. Let us mention that in case the pairs of nodes in E are ordered, the graph is called directed, but we do not need to consider directed graphs in the networks context here.

**Definition**: The order of a graph is the number of its vertices, its size is the number of its edges, and the degree of a node is the number of nodes connected to it. The nodes with high degrees are called hubs.

**Remark**: We limit ourselves to graphs without loops, but in some cases, such limitation might be counterproductive.

**Definition**: Scale-free networks are distinguished by the property that their degree distribution follows a power-law relationship:
$$P(k) \sim k^{-\gamma},$$
where the probability $P(k)$ that a node in the network connects with k other nodes is roughly proportional to $k^{-\gamma}$, and this function gives a good fit to the observed data. The coefficients $\gamma$ vary for most of the networks from 2 to 3.